\documentclass{article}
\usepackage{amsmath}
\usepackage{amsfonts}

\newcommand{\di}{\displaystyle}

\newcommand{\R}{\mathbb  R}

\newcommand{\N}{\mathbb  N}
\newcommand{\1}{\mathbb  I}
\newcommand{\la}{\langle}
\newcommand{\ra}{\rangle}

\newtheorem{lemma}{Lemma}[section]
\newtheorem{theorem}[lemma]{Theorem}
\newtheorem{proposition}[lemma]{Proposition}
\newtheorem{corollary}[lemma]{Corollary}
\newtheorem{remark}[lemma]{Remark}

\newtheorem{definition}[lemma]{Definition}

\newtheorem{hypothesis}[lemma]{Hypothesis}

\def\noi{\noindent}

\def\beq{\begin{equation}}   \def\eeq{\end{equation}}
\def\bea{\begin{eqnarray}}  \def\eea{\end{eqnarray}}
 
\def\noi{\noindent}

\newcommand\mysection{\setcounter{equation}{0}\section}
\renewcommand{\theequation}{\thesection.\arabic{equation}}
\newcounter{hran} \renewcommand{\thehran}{\thesection.\arabic{hran}}

\def\bmini{\setcounter{hran}{\value{equation}}
    \refstepcounter{hran}\setcounter{equation}{0}
    \renewcommand{\theequation}{\thehran\alph{equation}}\begin{eqnarray}}

\def\bminiG#1{\setcounter{hran}{\value{equation}}
\refstepcounter{hran}\setcounter{equation}{-1}
\renewcommand{\theequation}{\thehran\alph{equation}}
\refstepcounter{equation}\label{#1}\begin{eqnarray}}

\begin{document}

\title {Irregular time dependent perturbations of quantum Hamiltonians
\footnotemark\footnotetext{This work  was supported by the French Agence Nationale de la Recherche,
 NOVESOL project, ANR 2011, BS0101901}}
\author{ Didier Robert \\
%D\'epartement de Math\'ematiques\\
%Laboratoire Jean Leray, CNRS-UMR 6629\\
 %Universit\'e de Nantes, 2 rue de la Houssini\`ere, \\
%F-44322 NANTES Cedex 03, France\\
%{\it didier.robert@univ-nantes.fr}
%   in  memory  of Yuri Safarov 
}
\vskip 1 truecm
\date{}
\maketitle

\begin{abstract}
Our  main goal  in this paper is to prove existence (and uniqueness)  of the quantum  propagator  for time dependent quantum Hamiltonians  $\hat H(t)$ when  this  Hamiltonian  is perturbed with a   quadratic white noise
 $\dot{\beta}\hat K$.   $\beta$ is a continuous  function in time $t$, $\dot \beta$ its time derivative and   $K$ is a quadratic Hamiltonian. $\hat K$ is the Weyl quantization of $K$. \\ 
For time dependent  quadratic Hamiltonians $H(t)$   we recover, under less restrictive assumptions,  the   results  obtained  in \cite{ bofu, du}.
In our  approach we use  an exact Hermann Kluk formula \cite{ro2}  to deduce a  Strichartz    estimate for the propagator of $\hat H(t) +\dot \beta K$.  \\
 This is  applied to obtain local and global well posedness  for solutions for non linear Schr\"odinger equations with   an irregular  time dependent linear part.

\end{abstract}

\pagestyle{myheadings}

\mysection{Introduction }
The  linear  time  dependent  Schr\"odinger equation was  studied  in \cite{ya1, ya2}  for  time  dependent potential at least  continuous  in time. There are  physical  motivations  to  consider    Schr\"odinger  equations perturbed  by quadratic  potentials times  a white  noise  in time  (see  for    \cite{bofu, du}).   But  the  constructions  elaborated  in   Fujiwara \cite{fu1} and Yajima \cite{ya1}   are  no more  valid for   time discontinuous Hamiltonians.  
Nevertheless  it has been  shown in  
\cite{bofu, du} that  these  constructions,   based  on   Fourier  integral representations of the  propagator,  can  be revisited  and extended when the  time  dependence in the  Hamiltonian  is irregular, as  for  white noise.  The  main motivation for  considering  irregular
 time  dependent  Hamiltonian comes  from  Bose Einstein  condensation   or  fiber optics  (see \cite{bofu}  and its  bibliography).\\
The main   idea  developped in \cite{du}  to construct a propagator  $U_\beta(t,s)$ for  the  quantum hamiltonian 
    $\hat H(t) + \dot{\beta}\hat K$, where $\beta$ is in some H\"older class (so its derivative $\dot\beta$  is a distribution in time),   is  to establish  a suitable  representation formula  when  $\beta$  is $C^1$-smooth in time    and to prove that
    the  dependence   of $U_\beta(t,s)$ in  $\beta$  is   continuous for  the  topology  of the  H\"older  space $C^\mu(I_T)$, $0\leq\mu<1$  in a time interval $I_T=[t_0-T, t_0+T]$, $T>0$.\\
    This  strategy  was initiated  by Sussmann \cite{su}  for  solving  stochastic differential  equations by  deterministic methods.
 
 In this paper we shall  extend  the main results  of \cite{bofu, du}  to  more  general  quadratic  hamiltonians by  using a different  approach. Instead  of  establishing  a generalized  Mehler  formula  for  the  time dependent  propagator  we  choose to  use  a formula
  inspired from  the Hermann-Kluk formula \cite{ro2}, which  is more  flexible. The advantage  of this approach is that  the  link between  classical and quantum  mechanics  is straightforward, we do not need to  take care of caustics  because it is not  necessary  to solve the classical Hamilton-Jacobi equation as in the H\"ormander-Maslov approach.   The quantum  oscillations  are  represented  by  complex phases  so that  a Melher  (or Van Vleck)  type  formula  can be  recovered  by a  stationary phase  argument.  
  This is related to  complex WKB analysis and   coherent  states (see \cite{coro1} and its  bibliography).\\
 We shall  extend here  several well  known results in the ${\cal C}^1$ time  regular case  when $\beta$  in only  continuous or in an  H\"olderian  class of order $\mu\in]0, 1[$. For example  $\beta$ could  be a trajectory  of a one dimensional Brownian motion ($\mu =\frac{1}{2}-\varepsilon$)
     or of  a  fractional  Brownian motion ($\mu = H-\varepsilon$, $H\in]0, 1[$ being the Hurst index of the process).
 
 %%%%%%%%%%%%%%%%
 %%%%%%%%%%%%%%%
 \noi
 {\bf Acknowledgements}.   The author thank Laurent Thomann for his  comments on a preliminary version of this paper 
 %  and the referee   for the relevant  suggestions.

 \section{Mathematical Settings  and Results   }
Let   $H(t)$   be  a time dependent real  polynomial  Hamiltonian, of degree  at most 2 in the phase space space $\R^d_q\times\R^d_p$,  with  continuous coefficients in $t\in I_T=[t_0-T, t_0+T]$, $K$  is  a real  polynomial time independent Hamiltonian of degree  at most 2.\\
  Let $\beta$ be  a continuous function  of time $t$.   Denote $H_\beta(t) = H(t) + \dot \beta K$
an   irregular perburbation  of $H(t)$  where $\dot w =\frac{d\beta}{dt}$. $H_\beta$ is here  a distribution  in the  time $t$  so  the  meaning  of the classical Hamilton equation is not  clear. Denote $z=(q,p)$ a generic point  in the phase  space $\R^d_q\times\R^d_p$.  The  classical  Hamilton system is:

\beq\label{ham}
\begin{cases}
\dot q_\beta &= \partial_p H_\beta(q_\beta, p_\beta) \\
\dot p_\beta &= -\partial_q H_\beta(q_\beta, p_\beta)\\
\end{cases}
\eeq
 where $q_\beta=q_\beta(t,s)$, $p_\beta=p_\beta(t,s)$ with initial data  at time $t=s$,  $q_\beta(s,s)=q(s)$, $p_\beta(s,s)=p(s)$.\\
If $\beta$  is $C^1$  the  classical evolution (\ref{ham})  is linear and    so well  defined. 
Let us denote $\Phi_\beta(t,s)z = (q_\beta(t,s),p_\beta(t,s))$ where $z=(q,p)$, $q=q(s)$, $p=p(s)$.  Using   the  method 
developed in \cite{su, du} the Hamiltonian flow $\Phi_\beta(t,s)$ can  be extended  in a natural  way as a symplectic map  for $\beta\in{\cal C}^0$.\\
Let us now consider  the quantum  evolution.  If  $\beta$ is $C^1(I_T)$ denote by 
$\hat H_\beta(t)$ the Weyl  quantization of $H_\beta(t,q,p)$ (see for example \cite{coro1}). Here  the Planck constant is fixed, so we choose $\hbar =1$.\\
It is well known that $\hat H_\beta(t)$  is a self-adjoint  operator  in $L^2(\R^d)$  and that  the 
 time dependent Schr\"odinger equation  generates  a  continuous family of  unitary operators  in $L^2(\R^d)$,  which we denote by $U_\beta(t,s)$,   satisfying
 \beq
 i\partial_t U_\beta(t,s) = \hat H_\beta(t)U_\beta(t,s),\;\; U_\beta(s,s)=\1.
 \eeq
Denote $C_R^0(I_T)=\{\beta\in C^0(I_T), \Vert \beta\Vert_\infty \leq R\}$  and  
$C^1_R(I_T) = C_R^0(I_T)\bigcap C^1(I_T)$ (equipped  with  the sup-norm).
We have the following preliminary result.
\begin{theorem}\label{main1}
(i) The map 
 $\beta\mapsto \Phi_\beta$  is a Lipschitzian  map  from $C_R^1(I_T)$
 into $C^0(I_T\times I_T , S(2d))$  where $S(2d)$ is  the space of linear  symplectic maps  of $\R^{2d}$.
In particular for any $T>0$, the map  $(\beta, t, s)\mapsto\Phi_\beta(t,s)$ can  be extended 
 in a unique continuous map from   $C_R^0(I_T)\times I_T\times I_T$  into  the affine symplectic group  of $\R^{2d}$. \\
 (ii) For  any $\psi\in{\cal S}(\R^d)$, the map $\beta\mapsto U_\beta(t,s)\psi$ is uniformly  continuous on $C_R^1(I_T)$. 
In particular  the map $(\beta, t, s)\mapsto U_\beta(t,s)$   can  be extended 
 in a unique continuous map from   $C_R^0(I_T)\times I_T\times I_T$  into  the unitary  group of $L^2(\R^d)$.
\end{theorem}
Part (i) of Theorem \ref{main1}  will be proved in section 3 and  part (ii)  in section 4.2. \\
Let  ${\cal K}_\beta(t,s;x,y)$ be  the (distribution )  Schwartz  kernel  of $U_\beta(t,s)$.
The next  result  is an exact   formula for ${\cal K}_\beta(t,s;x,y)$  depending only on  the  classical  dynamics $\Phi_\beta(t,s)$. A more  precise statement will  be given later  in Corollary \ref{pert1}.
\begin{theorem}\label{main2}
For every $\beta\in{\cal C}^0(I_T)$ there  exist  explicit  complex functions  $a_\beta(t,s)$  and $\Psi_\beta(t,s,z, x,y)$  where  
 $t,s\in I_T$, $z=(q,p)\in\R^{2d}$, $x,y\in\R^d$, such that
 \beq\label{kern0}
 {\cal K}_\beta(t,s;x,y) = a_\beta(t,s)\int_{\R^{2d}}\exp\Big(i\Psi_\beta(t,s,z,x,y)\Big)dz
 \eeq
Moreover $\Psi_\beta$  is polynomial of degree at most 2 in $z$  and $ \Im\Psi_\beta \geq 0$.\\
$(\beta, t, s)\mapsto a_\beta(t,s)$  and $(\beta, t, s)\mapsto \Psi_\beta(t,s,z,x,y)$  are  continuous  on  $C_R^0(I_T)\times I_T\times I_T$.\\
The equality  (\ref{kern0})  is an equality  between two distributions in the Schwartz space  ${\cal  S}^\prime(\R^d_x\times\R^d_y)$.
\end{theorem}
Notice that   the formula (\ref{kern0}) is valid without  condition on the time interval $I_T$, the caustics are not obstructions here. Of course this difficulty appears  again   when computing   the integral  in $z\in\R^{2d}$ to get the following result. 
To go further we need the following hypothesis.
\begin{hypothesis}\label{MVV1} The Hessian matrix $\partial_{p,p}^2 H$ (constant  here)  is  non   singular   and  \\$\partial_{p,p}^2 K = 0$. 
\end{hypothesis}
\begin{hypothesis}\label{MV2}   $\partial_{q,p}^2 K=0$  or $\beta\in{\cal C}^\mu(I_T)$  with $\mu>\frac{1}{2}$.
\end{hypothesis}
    \begin{remark}
If $d=3$   and if $\partial_{q,p}^2 K$  is an antisymmetric matrix,  it represents   an angular momentum  rotation term. This case was considered in \cite{ams}   without  noise
 and in $\cite{du}$  for perturbations   by noise. 
    \end{remark}
    The following  result  is a consequence of section 4.1 and section 5.1.
 \begin{theorem}\label{main3} Assume that  Hypothesis \ref{MVV1}   and Hypothesis \ref{MV2}  are satisfied. \\
    (I) For  every $R>0$ there  exists $T_R>0$  such that for  every $t,s\in I_{T_R}$
      and every $\beta$ such that if $\beta\in{\cal C}^\mu(I_T)$, $\Vert \beta\Vert_{\cal C^\mu}\leq R$ and $t\neq s$, 
     the Schwartz kernel ${\mathcal K}_\beta(t,s)$ of   $U_\beta(t,s)$   is a $C^\infty$  function of $(x,y)$  given by the following formula
     \beq\label{MVV}
     {\mathcal K}_\beta(t,s;x,y) = b_\beta(t,s)^{-d/2}{\rm e}^{iS_\beta(t,s;x,y)}
    \eeq
      where  $b_\beta(t,s)$ is continuous in $(\beta, t, s)$, $t\neq s$,  $S_\beta(t,s;x,y)$  is the classical action  along  the unique  classical  trajectory  joining
      $y$ to $x$ at time $s$.\\
      Morever there exists $\gamma >0$ such that 
       $\vert b_\beta(t,s)\vert \geq \gamma\vert t-s\vert$  for  every $t,s\in{I_T}_R$. \\
(II) There  exists  a constant $C_R$, depending  only on $R$ such that  for  every $t,s\in{I_T}_R$   and every $x, y\in\R^d$, we have 
  \beq\label{unifest}
   \vert{\mathcal K}_\beta(t,s;x,y)\vert \leq C_R\vert t-s\vert^{-d/2}.
  \eeq
  and for every $p\in[2, +\infty]$,   we have  for $\psi\in {L^p(\R^d)}$, 
  \beq\label{stri0}
  \Vert U_\beta(t,s)\psi\Vert_{L^p(\R^d)}  \leq 
  C_R\vert t-s\vert^{-d(1/2-1/p)}\Vert\psi\Vert_{L^{p'}(\R^d)},\; 1/p+1/p' = 1.
  \eeq
\end{theorem}
As it is well known the dispersive   estimate (\ref{stri0})    is closely related with Strichartz estimates (see \cite{keta}) and allows application  to non linear 
 Schr\"odinger  equations.  The case with noise was considered in \cite{bofu, du}.\\
   Let us consider the non linear Schr\"odinger equation (NLS):
 \beq\label{NLSD}
 i\partial_t\psi = \hat H_\beta(t)\psi + \lambda\vert\psi\vert^{2\sigma}\psi,\;\psi(s)=\psi, 
  \eeq
  where $\lambda\in\R, \sigma>0$. \\
       Here $H_\beta(t)$  is  irregular  in time $t$ so we have to  consider 
   the following integral mild  version of (\ref{NLSD})
    \beq\label{NLSI}
\psi(t)  = U_\beta(t,s)\psi -i \lambda\int_{s}^tU_\beta(t,u)\vert\psi(u)\vert^{2\sigma}\psi(s)du.
  \eeq
Let us introduce the Sobolev weighted spaces  associated with the harmonic oscillator:
$$
{\cal H}^k(\R^d) = \{\psi\in L^2(\R^d),\;\; \psi\in H^k(\R^d),\; \vert x\vert^k\psi\in L^2(\R^d)\}
$$
where  $H^k(\R^d)$, $k\in\N$,   is the usual Hilbertian Sobolev  space.\\
We shall see that these  spaces  are  invariant by the  quantum  propagator $U_\beta(t,s)$
 for  any $\beta\in{\cal C}^0(I_T)$. In order    to  include  the Gross-Pitaevski non linearity ($\sigma = 1$) we have to  consider initial data  in the  space ${\cal H}^1(\R^d)$.  Here  we have  the following  local result  proved in section 5.
 \begin{theorem}\label{main4}
 We assume that  Hypothesis \ref{MVV1}  and Hypothesis \ref{MV2}  are  satisfied.\\  
(I) If  $0<\sigma< \frac{2}{d}$, then then for any 
$\psi\in L^2(\R^d)$ the  integral equation (\ref{NLSI}) has a unique solution 
$\psi_\beta\in {\mathcal C}^0(I_T,L^2(\R^d))$. Moreover,  for every $T>0$,  we have 
$\psi_\beta\in {\cal C}^0\Big(I_T,L^2(\R^d)\Big)\cap  L^r\Big(I_T, L^{2\sigma}(\R^d)\Big)$,   with 
$r = \frac{4(\sigma+1)}{d\sigma}$.  \\ The $L^2$ norm  is conserved:
 $\Vert\psi_\beta(t)\Vert_{L^2(\R^d)} = \Vert\psi\Vert_{L^2(\R^d)}$ for every $t\in I_T$.\\
 Moreover if  $\psi\in {\cal H}^1(\R^d)$  then $\psi_\beta\in {\mathcal C}^0(\R,{\cal H}^1(\R^d))$.\\
 (II) If  $0<\sigma< \frac{2}{d-2}$, $d\geq 3$. Then for any 
$\psi\in {\cal H}^1(\R^d)$   there exists $0<T=T(\Vert\psi\Vert_{{\cal H}^1(\R^d)},s)$ such that
the  integral equation (\ref{NLSI}) has a unique solution 
$\psi_\beta\in {\mathcal C}^0(I_T, L^2(\R^d))\cap  L^r\Big(I_T, L^{2\sigma}(\R^d)\Big)$,   with 
$r = \frac{4(\sigma+1)}{d\sigma}$  and such that  for any $a, b\in\R^d$, 
 $$
 (a\cdot x + b\cdot\nabla_x)\psi_\beta\in {\mathcal C}^0(I_T,L^2(\R^d))\cap  L^r\Big((I_T), L^{2\sigma}(\R^d)\Big).
 $$
\end{theorem}
\begin{remark}
A global well-posedness result in the ${\cal H}^1$-subcritical   case with a rotation term  for (\ref{NLSI})  is proved in \cite{ams} for time independent Hamiltonians. 
For $\sigma<\frac{2}{d}$ ($L^2$-subcritical  non linearity )  a global result  in 
${\cal H}^1(\R^d)$ could be obtained under the assumptions of Theorem \ref{main4} following \cite[Theorem 2.2]{ams} 
  but for    $\frac{2}{d} \leq \sigma <\frac{2}{d-2}$, $d\geq 2$  and $\lambda \geq 0$ (defocusing case)   the situation is much more involved because  we cannot use  the energy conservation. In \cite[Theorem 3]{du}  the author uses the result  of  \cite{ams}
to get a global  ${\cal H}^1$-well-posedness result with a rotation term  in the regular part H  assuming  that  $K$ is linear  in $(q,p)$.\\
Notice that global well-posedness results in the  supercritical case are  proved in \cite{PRT}   for the Gross-Pitaevski equation  with  random  initial data.
    \end{remark}
    \begin{remark}
   The  Non linear Schr\"odinger equation  with   a white noise dispersion is   also considered in \cite{bode}  (see also  the references of \cite{bode}). These papers    use  the   probabilistic setting  of stochastic processes.
    \end{remark} 
%%%%%%%%%%%%%%%%%
%%%%%%%%%%%%%%%%%

\section{The  smooth time  dependent case}

In this  section  we assume  that  $H(t)$   is a time dependent  polynomial  Hamiltonian, of degree  at most 2,  with  continuous coefficients in $t\in I_T=[t_0-T, t_0+T]$, $K$  is  time independent.
  Let $\beta$  be a continuous function  of time $t$.   Denote $H_\beta(t) = H(t) + \dot \beta K$
 the  irregular perburbation  of $H(t)$  where $\dot\beta =\frac{d\beta}{dt}$. \\
If $\beta$  is $C^1$ the  classical  and quantum  evolutions  are  well  defined.  We   shall   show in the next section   that  these evolutions  are  still  well defined  for $\beta\in C^0(I_T)$  following  an approach inspired  from \cite{su,du}.\\
We  shall review  here  some more or less well known formulas concerning  quantum time dependent quadratic Hamilonians.\\
It is  well known that  classical   and quantum  evolution  are  well defined  if  $H(t)$ is continuous in time (and for $H_\beta(t)$ if $\beta$ is $C^1$) and there  exist exact formulas  related  the classical  and quantum evolution.
 Denote  by $\Phi_{H}(t,s)$  the  classical   flow in the phase space $\R^{2d}$, at  time $t$  with initial data at $s$ and  by $U_{H}(t,s )$ the
 quantum  propagator generated by  the Weyl  quantization $\hat H(t)$ of $H(t)$. \\
  Let  us recall    now  a formulation of  the  exact    correspondence   classical-quantum.
We  have  $H(t) = H_2(t) + H_1(t)+ H_0(t)$  where  $H_j(t)$ is  a polynomial  of degree $j$ in $(q,p)\in\R^{2d}$.  Let us denote $S_{H_2}(t)$ the   matrix of the quadratic form $H_2(t)$. More  explicitly: 
      $$
   H_2(t;q,p) = \frac{1}{2}\left(G_H(t)q\cdot q + 2L_H(t)q\cdot p + E_H(t)p\cdot p\right)
   $$ 
   where $(q,p)\in\R^d\times\R^d$, 
    $S_{H_2}(t) = \begin{pmatrix}G_H(t) &L_H(t)^\top\\L_H(t) &E_t\end{pmatrix}$, 
   $q, p \in\R^d$,  $G_H(t), L_H(t), E_H(t)$ are real,  $d\times d$  matrices,
   continuous in time $t\in\R$, $E_H(t), G_H(t)$ are symmetric, $L_H(t)^\top$  is the  transposed matrix of $L_H(t)$. The classical motion defined by $H_2(t)$ in the phase space $\R^{2d}$, 
    is given by the linear differential equation
    \beq\label{classicevol}
    \left(\begin{array}{c}\dot q\\ \dot p\end{array}\right) 
     = J.\left(\begin{array}{cc}G_H(t) & L_H(t)^T\\
     L_H(t) & E_H(t) \end{array}\right)\left(\begin{array}{c}q \\p\end{array}\right),\;\;\;J = \begin{pmatrix} 0&\1\\-\1&0\end{pmatrix}
     \eeq
    where the matrix  $J$  defines the symplectic form $\sigma(z,z^\prime) := Jz\cdot z^\prime$,
      $z=(q,p)$, $z^\prime=(q^\prime,p^\prime)$, $z\cdot z^\prime$ denotes the scalar product in $\R^{2d}$. \\
     This equation defines a linear symplectic transformation, $\Phi_{H_2}(t,s)$, 
      such that  $\Phi_{H_2}(s,s) = \1$.    It can be represented        
       as a     $2d\times 2d$ matrix  
  which can be written as four $d\times d$ blocks~:
 \beq 
\Phi_{H_2}(t,s)= \left(\begin{array}{cc}A(t,s) & B(t,s) \\
C(t,s)&  D(t,s) \end{array}\right).
\eeq
Let  us denote $\hat H$  the Weyl  quantization    of   the Hamiltonian H (see \cite{coro1}  for 
the  definition  and properties of the Weyl quantization). \\
Let  us denote ${\cal K}_{H_2}(t,t_0;x,y)$  the Schwartz  kernel  of the  quantum  propagator $U_{H_2}(t,t_0)$.  
\\ It is known  that  the propagator   $U_{H}(t,t_0)$  is well  defined   (\cite{coro1}, p.67).  
  It is  unique and satisfies the following  properties.  
      \beq
  i\partial_t U_H(t,s) = \hat H(t)U_H(t,s);\;\;  U_H(s,s)=\1
  \eeq
    Let  ${\cal K}_H(t,t_0)\in{\cal S}^\prime(\R^d_x\times\R^d_y)$  be the  Schwartz Kernel  of
     $U_H(t,t_0)$. There  exist    many papers giving more  or less explicit  formula  for 
     ${\cal K}_H(t,t_0)$ (\cite{ho, ni, ro2}). For our purpose it is convenient  to use 
      a  formula  closely related with  coherent states and   symplectic geometry  of the phase  space (for details  see \cite{ro2}).\\
      Let us introduce  the Siegel space  $\Sigma_+(d)$ of $d\times d$  complex matrices $\Gamma$  with imaginary part $\Im \Gamma :=\frac{\Gamma-\Gamma^\top}{2i}$
       definite-positive. Let be $\Theta$ be a continuous map from $I_T$ into  $\Sigma_+(d)$   and $M_\Theta(t,t_0) = (C(t,t_0)-iD(t,t_0) -\Theta(t)\left(A(t,t_0)-iB(t,t_0)\right)$.
     The exact  correspondence between  classical  and  quantum mechanics can be expressed as follows.   Let us denote ${\cal K}(t,t_0;x,y)$ the  Schwartz kernel of the quantum propagator $U_{H_2}(t,s)$.
  \begin{proposition}[Hermann-Kluk formula in the quadratic case \cite{ro2}]
 We have the following exact formula
 \beq\label{kquad}
 {\mathcal K}_{H_2}(t,t_0;x,y) = 2^{d/2}(2\pi)^{-3d/2}{\rm det}^{-1/2}\Bigl(\frac{M_\Theta(t,t_0)}{i}\Bigr)\int_{\R^{2d}}
 {\rm e}^{i\Psi_{\Theta,2}(t,t_0;z;x,y)}dz
 \eeq
 where 
\bea
\Psi_{\Theta,2}(t,t_0;z;x,y) = \frac{1}{2}(q_t\cdot p_t -q\cdot p) + p_t\cdot(x-q_t) -p\cdot(y-q) 
 \nonumber  \\+\frac{1}{2}
\bigl(\Theta(t)(x-q_t)\cdot(x-q_t) + i(y -q)\cdot(y-q)\bigr),
\eea
$z=(q,p)\in\R^d\times\R^d$.
  \end{proposition}
\begin{remark}
$\Theta$  is a useful  degree of freedom to compute  ${\mathcal K}_{H_2}(t,t_0;x,y)$.
The choices $\Theta = i\1$    and $\Theta(t) = \Gamma(t,t_0)$ can be useful,  where\\
 $\Gamma(t,t_0) = C(t,t_0)+iD(t,t_0)(A(t,t_0)+iB(t,t_0))^{-1}$.\\
 In \cite{ro2} $\Theta$ is supposed to be $C^1$  in $t$. The result  is clearly valid for $\Theta$ only continuous.  In formula (2.12) of \cite{ro2} we have to read $\det^{-1/2}$  and not $\det^{1/2}$.  Notice that   $M_\Theta(t,t_0)$ is  invertible (property  of the action of symplectic matrices on the  Siegel space).
   
\end{remark}
 
  Adding  now   lower order terms  we get
  \begin{corollary}\label{pert1}
Suppose now  that  $H(t) = H_2(t) + H_1(t)+H_0(t)$  where  $H_j(t)$ is homogeneous of degree $j$ in
 $z=(q,p)\in\R^{2d}$.\\
Then the  Schwartz kernel ${\mathcal K}_{H}(t,t_0;x,y) $  of $U_H(t,t_0)$   has the following  expression
\beq\label{kpol}
 {\mathcal K}_{H}(t,t_0;x,y) = 2^{d/2}(2\pi)^{-3d/2}{\rm det}^{-1/2}\Bigl(\frac{M(t,t_0)}{i}\Bigr)\int_{\R^{2d}}
 {\rm e}^{{i}\Psi(t,t_0;z;x,y)}dz
 \eeq
where 
$$
\Psi_\Theta(t,t_0,z,x,y) = \Psi_{\Theta,2}\left(t,t_0;z;x,y +\int_{t_0}^t\tilde{b}(s)ds\right)-\int_{t_0}^t\tilde{a}(s)y_sds -\int_{t_0}^t H_0(s)ds
$$
with $y_s =y + \int_t^s\tilde b(s)ds$, $\tilde a, \tilde b$ depend on $H_1(t)$ and are  given  in the  proof.\\
When applied to $H_\beta(t)$ (here $\beta\in{\cal C}^1(I_T)$) we use the notations ${\cal K}_\beta={\cal K}_{H_\beta}$
 and $\psi_\beta = \psi$.
\end{corollary}
{\bf Proof}.  It is enough to assume that $H_0(t)=0$. Recall here  a well  known argument (Lagrange  method).  Let  us compute $V(t,t_0)$   such that
  $U_H(t,t_0) = U_{H_2}(t,t_0)\cdot V(t,t_0)$.  We  get  the  following  equation:
   \beq\label{lag}
 i\partial_tV(t,t_0) = U_{H_2}(t_0,t)H_1(t)U_{H_2}(t,t_0)V(t,t_0).
 \eeq
  We have $H_1(t;q,p) = a(t)\cdot q + b(t)\cdot p$. Using  the exact  Egorov formula  \cite{coro1} for  quadratic Hamiltonians we get
  $$
  U_{H_2}(t_0,t)\hat H_1(t)U_{H_2} = \hat A(t),
   $$
  where $A(t,z) = H_1(t,\Phi_{H_2}(t,t_0)z)$. 
Then by the  characteristics method we get 
  $$
  V(t,t_0)\psi(t_0, x) = \exp\left({i}\int_{t_0}^t\tilde{a}(s)x_sds\right)\psi\left(t_0, x-\int_{t_0}^t\tilde{b}(s)ds\right)
  $$
  where 
  $\begin{pmatrix} \tilde a \\  \tilde b \end{pmatrix}= \Phi_{H_2}(t,t_0)^\top\begin{pmatrix} a \\  b \end{pmatrix}$,  $N^\top$  is  the transposed matrix of the  matrix $N$, 
  $x_s = x+\int_t^s\tilde b(s)ds$.
The corollary follows.\\

$\square$

 %%%%%%%%%%%%%%
 %%%%%%%%%%

\section{The time irregular case}
\subsection{The classical evolution}
What remains   true of the  previous  computations for $H_\beta(t) = H(t) + \dot \beta K$  when  $\beta$  in only continuous  in $I_T$?\\
For the noise part $\dot \beta K$ the classical evolution  is  linear : $\Phi_{\dot \beta K}(t,s,z) = \Phi_K(\beta_t-\beta_s)z$. Let $z(s)\in\R^{2d}$  be an initial data. Then $z_\beta(t)= \Phi_K(\beta_t,\beta_s, z_s) $  is  a   solution in  Sussmann \cite{su}  sense of the Hamilton equation  
\beq\label{ham1}
\dot z_\beta(t) = \dot \beta(t)J\nabla_z K(z_\beta(t)),\;\;z_\beta(s)=z(s).
\eeq
We have  here  
$$
 z_\beta(t) =\exp\Big((\beta(t)-\beta(s))JS_K\Big)z
$$
%The  flow  $z\mapsto z_\beta(t)$ clearly  satisfies the   definition   given below, considered by 
% Sussmann when $\beta$ is  continous  but not differentiable.

Now let us consider 	the   perturbed  Hamiltonian $H_\beta(t) = H(t) + \dot \beta K(t)$. We want to  define  a classical trajectory
   $z_\beta(t) = \Phi_{H_\beta}(t,s,z(s))$  for the perturbed Hamilton  equation
   \beq\label{ham2}
\dot z_\beta(t) = J\nabla_z H_\beta(z_\beta(t)),\;\;z_\beta(s)=z(s).
\eeq

%Let us  denote $\beta^\varepsilon (t) = \frac{1}{2\varepsilon}\int_{t-\varepsilon}^{t+\varepsilon} \beta(\tau)d\tau$.
\begin{definition} $z_\beta(t)$ is a Sussmann  solution of (\ref{ham2}) if
\begin{itemize}
\item[(CL0)] There  exists  a neighborhood ${\cal N}_\beta$ of $\beta$ in $C^0(I_T)$  such that if
 ${\cal N}_\beta^1 = {\cal N}_\beta\bigcap C^1(I_T)$ then $\tilde \beta\mapsto z_{\tilde \beta}(t)$ is  a uniformly continuous  map from 
 ${\cal N}_\beta^1$ into $C^0(I_T, \R^{2d})$.
\item[(CL1)] For every $\varepsilon >0$, $z_{\beta^\varepsilon}(t)$   solves (\ref{ham2})  for the $C^1$  function $\beta^\varepsilon$
\item[(CL2)] ${\di \lim_{\varepsilon\rightarrow 0}z_{\beta^\varepsilon}(t) = z_\beta(t) \;\;{\rm in}\;\; C^0(I_T,\R^d)}$. 
\end{itemize}
$C^0(I_T)$ is   equipped  with its natural norm  $\di{\Vert \beta\Vert_\infty = \sup_{t\in I_T}\vert \beta(t)\vert}$.\\
Properties (CL0), (CL1)  and (CL2)  define   a unique  mild solution   of (\ref{ham2}). In particular $z_\beta(t)$ is independent  on  the $C^1$ approximations  $\beta^\varepsilon$  of $\beta$. 
\end{definition}
We have  to prove that  conditions (CL0), (CL1)  and (CL2)  are  fulfilled.
 
Recall  that $H(t)$ and $K$  are  quadratic forms on $\R^{2d}$. 
If  $\beta\in C^1(I_T)$ then it is well known that $ \Phi_{H_\beta}(t,s)  $    is   a  symplectic linear transformation 
of  the  phase space $\R^{2d}$.  It  is convenient  here  to consider that  the Hamiltonian $H_\beta$  is a perturbation  of  the noise term $\dot \beta K$. Then  
 it  solves the following integral equation
\beq\label{hamint}
\Phi_{H_\beta}(t,s) = \Phi_K(\beta_t-\beta_s) + \int_s^t\Phi_K(\beta_t-\beta_\tau)JS_H(\tau)\Phi_{H_\beta}(\tau,s)d\tau,
\eeq
where $S_H(t)$ is  the  symmetric matrix of the quadratic form  $H(t)$.\\
Now the   trick  is that we can solve equation (\ref{hamint}) using the  Picard  fixed theorem.
Denote $C_R^0(I_T)=\{\beta\in C^0(I_T), \Vert \beta\Vert_\infty \leq R\}$  and  
$C^1_R(I_T) = C_R^0(I_T)\bigcap C^1(I_T)$ (equipped  with  the sup-norm)
\begin{proposition}[see also \cite{du}, proposition 2.29]
(1) There  exists $T_R >0$  small enough  such  that  for $T\leq T_R$  and $\beta\in C_R^0(I_T)$, 
the equation (\ref{hamint})    has  a unique  solution defined for 
$(t,s)\in I_T\times I_T$.\\
(2) $\beta\mapsto \Phi_{H_\beta}$  is a Lipschitzian  map  from $C_R^1(I_T)$
 into $C^0(I_T\times I_T , S(2d))$  where $S(2d)$ is  the space of linear  symplectic maps  of $\R^{2d}$.\\
(3) $\Phi_{H_\beta}(t,s)$    satisfies
\beq\label{chapkol}
\Phi_{H_\beta}(t,t_1) =\Phi_{H_\beta}(t,s)\Phi_{H_\beta}(s,t_1),\;\; \forall t, t_1, s \in I_T.
\eeq
In particular for any $T>0$ $\Phi_{H_\beta}(t,s)$ can  be extended to $I_T\times I_T$ in a unique way such that for every $z\in\R^{2d}$, $z_\beta(t) = \Phi_{H_\beta}(t,s)z$ satisfies (CL0), (CL1) and (CL2).\\
\end{proposition}
{\bf Proof}
(1)  is a  direct application   of the  Picard  fixed point theorem. First from a  well known estimate  for linear ODE   we have, for some $\Gamma >0$, 
\beq\label{expest}
\Vert\Phi_K(\beta_t-\beta_s)\Vert \leq {\rm e}^{\Gamma\vert \beta_t-\beta_s\vert}.
\eeq
For $X\in C^0(I_T\times I_T , S(2d))$  denote
\beq\label{inteq}
F_\beta(X) = \Phi_K(\beta_t-\beta_s) + \int_s^t\Phi_K(\beta_t-\beta_\tau)JS_H(\tau)X(\tau,s)d\tau.
\eeq
So if $\beta\in C_R^0(I_T)$,  $F_\beta$  has unique  fixed point  
$X_\beta$ in $C^0(I_T \times I_T , S(2d))$
 for $T\leq T_R$. Moreover there  exists  $C>0$  such  that 
 $$
 \Vert X_\beta(t,s)\Vert \leq C {\rm e}^{2\Gamma R}
 $$
 and if $\beta\in C^1_R(I_T)$ then $X_\beta = \Phi_{H_\beta}$.\\
(2) $\beta\mapsto \Phi_K(\beta_t, \beta_s)$ is  $C^1$
 from  $C_R^0(I_T)$ into $S(2d)$. Choosing $T_R>0$ small enough, the derivative $D_X F_W(X_\beta)$ satisfies $\Vert D_X F_W(X_\beta)\Vert \leq \frac{1}{2}$. 
 Applying the  implicit  function theorem we get that 
 $\beta\mapsto \Phi_{H_\beta}$  is also $C^1$.\\
(3) is  now  easy to prove using that it is true  for $\beta\in C^1$.
$\square$

%%%%%%%%%%%%%
%%%%%   fin modiff w  %%%%%%%%
%%%%%%%%%

We can now add  the contribution  of order one. \\
 We have $H_\beta(t) = H_2(t)+H_1(t) + \dot \beta(K_2 + K_1)$.\\
  Denote
  $H_{\beta,2}(t) = H_2(t) + \dot \beta K_2(t)$, $H_{\beta,1}(t) = H_1(t) + \dot \beta K_1$.\\
  $H_{\beta,1}(t,z) = \left(V_H(t)+\dot \beta V_K\right)\cdot z$.\\ We have, using the Duhamel formula, for every $z\in\R^{2d}$, 
\beq\label{O1}
  \Phi_{H_\beta}(t,s)z = \Phi_{H_{\beta,2}}(t,s)z + 
  \int_{s}^t\Phi_{H_{\beta,2}}(t,u)J\left(V_H(u)+\dot \beta V_K\right)du
\eeq
  $\Phi_{H_{\beta,2}}(t,s)$  solves the integral  equation (\ref{hamint}).  So plugging
  (\ref{hamint}) for $H=H_2$   in (\ref{O1})  and  integrating  by parts   we get 
  \begin{corollary}\label{O2+O1}
  The map $\beta\mapsto   \Phi_{H_\beta}(t,s)z$ given by (\ref{O1})  is $C^1$  from \\
  $C^0(I_T)$ into $\R^{2d}$  and $z_\beta(t)$  satisfies  the properties (CL0), (CL1) and (CL2).\\
  In particular there  exists $C_R>0$  such that for all $z\in\R^{2d}$, $\beta_1, \beta_2\in C_R(I_T)$, we have 
  \beq\label{lipclass}
  \vert z_{\beta_1}(t) - z_{\beta_2}(t)\vert \leq C_R\Vert \beta_1-\beta_2\Vert_{\infty}\vert z\vert
  \eeq
  \end{corollary}
%%%%%%%%%%%%%%%%%%%%%%%%%
%%%%%%%%%%%%%%%%%%%%%%%%%
\subsection{The quantum  evolution}
Following \cite{du} we define the quantum evolution for the Hamiltonian 
$\hat H_\beta(t)$ when $\beta\in C^0(I_T)$  as follows. 
\begin{definition}
$t\mapsto\psi_\beta(t)\in L^2(\R^d)$, $t\in I_T $, is a mild solution of the Schr\"odinger  equation
\beq\label{schw}
i\partial_t\psi(t) = \hat H_\beta(t)\psi(t),\;\; \psi(t_0) = \psi_0.
\eeq
if  the   following  conditions are  satisfied
\begin{itemize}
\item[(QM0)] There  exists  a neighborhood ${\cal N}_\beta$ in $C^0(I_T)$  such that if
 ${\cal N}_\beta^1 = {\cal N}_\beta\bigcap C^1(I_T)$ then $\tilde \beta\mapsto \psi_{\tilde \beta}(t)$ is  a uniformly continuous  map from 
 ${\cal N}_\beta^1$ into $C^0(I_T, L^2(\R^d))$.
\item[(QM1)] For every $\varepsilon >0$, $\psi_{\beta^\varepsilon}(t)$ solves (\ref{schw})  for $\beta=\beta^\varepsilon$.\\
\item[(QM2)] $\di{\lim_{\varepsilon\rightarrow 0}\psi_{\beta^\varepsilon}(t) = \psi_{\beta}(t)}$ in 
$C^0(I_T)$.
\end{itemize}
Recall that $\beta^\epsilon$  are $C^1$approximations of $\beta$ in $C^0(I_T)$.  As in the classical case  $\psi_{\beta}(t)$ 
 is independent on the $C^1$ approximations $\beta^\varepsilon$  of $\beta$.
\end{definition}
For  simplicity  we shall  assume that that $H_\beta$ is  homogenous of degree  2. Adding  terms  of degree 1 and 0 is easy to check using  the Duhamel formula  as in corollary  \ref{O2+O1}. 
%%%%%%%%%%%%%%%%%%%%
As in the  time  regular case we  shall  show  now  that the  quantum evolution is    completely determined  by the  quantum  evolution  studied  above. In order  to  go   from  the regular  to the  irregular case  we use  the following  proposition. 

We use now the notation $U_\beta = U_{H_\beta}$.
\begin{proposition}\label{quantest2}
For any  $R>0$  there exists $C_R>0$ and $T_R>0$  such that for  all $\beta, \beta^{(1)}, \beta^{(0)}\in C_R^0({I_T}_R)\cap C^1({I_T}_R)$, all $\psi\in{\cal S}(\R^d)$  and $t,s\in{I_T}_R$, we have
\beq\label{quantlip1}
\Vert U_{\beta}(t,s)\psi \Vert_{L^2(\R^d)}  \leq  
C_R\Vert\psi\Vert_{L^2(\R^d)}
\eeq
\beq\label{quantlip2}
\Vert U_{\beta^{(1)}}(t,t_0)\psi - U_{\beta^{(0)}}(t,t_0)\psi\Vert_{L^2(\R^d)}  \leq 
C_R\Vert \beta^{(1)}-\beta^{(0)}\Vert_{\infty}\Vert\psi\Vert_{{\cal H}^2(\R^d)}
\eeq
\end{proposition}
\begin{corollary}\label{quadquant}
$\beta\mapsto U_\beta(t,s)$ can be extended in a unique way  to $C_R^0(I_T)$ such that properties 
(QM0), (QM1) and (QM2)  are   satisfied.\\
 Moreover  the Schwartz kernel of $U_\beta(t,s)$ is given by the  Hermann-Kluk formula (\ref{kquad})  with  the  generalized classical  flow $\Phi_{H_\beta}(t,s)$ determined by (\ref{O1}).
 Notice  that the  linear part  $\Phi_{H_{\beta,2}}(t,s)$  of the affine map  $\Phi_{H_\beta}(t,s)$  is  a  symplectic matrix
  denoted by $\Phi_{H_{\beta,2}}(t,s) = \begin{pmatrix}A_\beta(t,s) & B_\beta(t,s)\\ C_\beta(t,s) & D_\beta(t,s)   \end{pmatrix}$
\end{corollary}
\begin{corollary}\label{invH}
For every $k\geq 0$ we have $U_\beta(t,s){\cal H}^k(\R^d)\subseteq {\cal H}^k(\R^d)$. \\
Moreover there  exists $C_{R,t_0,T}>0$  such that  for $\psi\in{\cal H}^k(\R^d)$, $t,s\in I_T$, 
$\beta\in{\cal C}^0_R(I_T)$,  we have
\beq\label{invk}
\Vert U_\beta(t,s)\psi\Vert_{{\cal H}^k(\R^d)}\leq C_{R,t_0,T}\Vert\psi\Vert_{{\cal H}^k(\R^d)}.
\eeq
\end{corollary}
{\bf Proof}.  This a consequence  of (\ref{quantlip1})  and of the Egorov property : 
\beq\label{eg}
U_\beta(s, t)\hat AU_\beta(t,s) = \widehat{A\circ\Phi_\beta(t,s)}.
\eeq
We  start  by proving the corollary for $k=1$  hence by induction we get  the  result for any $k\in\N$.\\
Consider  the  linear  symbol $A(q,p) = a\cdot q +b\cdot p$  and let  $\psi\in{\cal H}^1(\R^d)$.
 Then using (\ref{eg}) and that $\Phi_\beta(t,t_0)$   is  an affine map  we get that 
 $$
 \hat A U_\beta(t,s)\psi = U_\beta(t,s)U_\beta(s,t) \hat A U_\beta(t,s)\psi\in L^2(\R^d).
 $$
So $U_\beta(t,s)\psi\in {\cal H}^1(\R^d)$  and (\ref{invk})  for $k=1$.
$\square$\\
We shall  use coherent  states with the  notations of \cite[chapter.1]{coro1} (here  we choose the Planck constant  $\hbar=1$).   The next two lemmas   will be used to prove Proposition \ref{kern1}
\begin{lemma}\label{unif}
There  exists $T_R>0$ small enough, $C_R>0$  such  that  for $t, s\in {I_T}_R$ and 
$\beta\in{\cal C}^1_R({I_T}_R)$  we have
\bea\label{kerbarg}
\sup_{Y\in\R^{2d}}\int_{\R^{2d}}\vert\la\varphi_Y, U_\beta(t,s)\varphi_X\ra\vert dX & \leq M_R \\
\sup_{X\in\R^{2d}}\int_{\R^{2d}}\vert\la\varphi_Y, U_\beta(t,s)\varphi_X\ra\vert dY & \leq M_R
\eea
\end{lemma}
{\bf Proof}.   
From \cite{coro2}, Proposition (5.7)   we  have
\beq\label{matco}
\la \varphi_{z+X}, U_\beta(t,s)\varphi_z\ra = a_\beta\exp\Big(-\big\vert z+\frac{X}{2}\big\vert^2 + \Lambda_\beta(z + \frac{X-iJX}{2})\cdot(z + \frac{X-iJX}{2})\Big)
\eeq
where $\Lambda_\beta = \big(\1+F_\beta)(\1+F_W+iJ(\1-F_\beta)\big)^{-1}$, $F_\beta = \Phi_{H_\beta}(t,s)$,
 $a_\beta = 2^d\det\big(\1+F_W+iJ(\1-F_\beta)\big)^{1/2}$. \\
Then using  \cite[Lemma 5.11]{coro2}  and (\ref{matco}), we  have
$$ 
\vert\la\varphi_Y, U_\beta(t,s)\varphi_0\ra\vert \leq
 \exp\left(-\frac{\vert Y\vert^2}{2(1+\lambda_\beta(t,s)}\right),
$$
where $\lambda_\beta(t,s)$ is the largest  eigenvalue of $\Phi_{H_\beta}(t,s)\Phi_{H_\beta}(t,s)^\top$.\\
But we have $\Vert\Phi_{H_\beta}(t,s)\Vert \leq C{\rm e}^{2\Gamma R}$. So for some $C_R>0$  we have
\beq\label{barg1}
\vert\la\varphi_Y, U_\beta(t,s)\varphi_0\ra\vert \leq {\rm e}^{-\frac{\vert Y\vert^2}{C_R}}
\eeq
 But  $U_\beta(t,s)$ is the metaplectic transformation  associated with $F_\beta$. More precisely, recall that we have (see \cite{coro2})  $U_\beta(t,s) = \hat R(F_\beta)$  and $\varphi_X=\hat T(X)\varphi_0$, where
  $\hat R$ denotes the metaplectic representation  and $\hat T$ the Weyl translation representation. In particular  we have the useful property
  \beq\label{repquant}
  \hat R(F_\beta)\hat T(z)\hat R(F_\beta)^* = \hat T(F_\beta z).
  \eeq
So  we get 
 \beq\label{xy}
 \la\varphi_Y, U_\beta(t,s)\varphi_X\ra = {\rm e}^{g(\beta)}\la\varphi_{Y-F_\beta X}, U_\beta(t,s)\varphi_0\ra,
 \eeq
 where $g(\beta) = \frac{\sigma}{2}(F_\beta X,Y)$, $\sigma(Z,Y) = JZ\cdot Y$ is the symplectic form. 
  From (\ref{repquant}), (\ref{barg1}) and (\ref{xy})   we get, $t,s\in{I_T}_R$,
\beq\label{barg2}
\vert\la\varphi_Y, U_\beta(t,s)\varphi_X\ra\vert \leq 
{\rm e}^{-\frac{\vert Y-\Phi_{H_\beta}(t,s)X\vert^2}{C_R}}
\eeq
Now choosing $T_R$  small  enough  we have $\Vert\Phi_{H_\beta}(t,s)^{-1}\Vert \leq 2$.
 Hence (\ref{kerbarg})  follows from (\ref{barg2}).  $\square$
 
We come now to a continuity propery of $U_\beta$ in $\beta$ (\ref{quantlip2}). For proving this property we shall use again coherent states. \\ Let  us denote $\delta U = U_{\beta^{(1)}}(t,s) - U_{\beta^{(0)}}(t,s)$.
 We have to establish an estimate  for the Bargman kernel $\tilde K_{\delta U}(X,Y):=\la\varphi_Y, \delta U\varphi_X\ra$.
 \begin{lemma}\label{delco}
 For any  $R>0$  there exists $C_R>0$  and $T_R$  such that for  all $ \beta^{(1)}, \beta^{(2)}\in C_R^0({I_T}_R)\cap C^1({I_T}_R)$, $X, Y\in\R^{2d}$, 
 \beq\label{coker}
 \vert\tilde K_{\delta U}(X,Y)\vert  \leq 
 C_R\Vert \beta^{(1)}-\beta^{(2)}\Vert_\infty(1+\vert F_{\beta^{(\theta)}}X\vert\vert Y\vert +\vert Y-F_{\beta^{(\theta)}}X\vert^2)
 {\rm e}^{-\frac{\vert F_{\beta^{(\theta)}}X-Y\vert^2}{C_R}}.
 \eeq 
 \end{lemma}
 {\bf Proof}.  We  use the same  method  as in  the proof of Lemma \ref{unif}. 
  For $\theta\in[0, 1]$  denote  $\beta^{(\theta)} = \theta \beta^{(1)} + (1-\theta)\beta^{(0)}$. 
  So  we   have 
  \beq\label{kdelta}
 \tilde K_{\delta U}(X,Y) = \int_0^1\frac{\partial}{\partial \theta}
 \la\varphi_Y, U_{\beta^{(\theta)}}\varphi_X\ra d\theta. 
\eeq
Using (\ref{matco}) and   known estimates  on $F_{\beta_\theta}$  we   shall easily get (\ref{coker}).  Let us begin with the particular case $X=0$. We have to compute 
$\frac{\partial}{\partial\theta} \la\varphi_Y, U_{\beta^{(\theta)}}\varphi_0\ra$   using (\ref{matco}). Then applying Corollary \ref{O2+O1}  and (\ref{barg1}) we get for every $\theta\in[0, 1]$, $C_R>0$ large enough, 
  \beq{\label{exp2}
  \vert\frac{\partial}{\partial\theta}\la\varphi_Y, U_{\beta^{(\theta)}}\varphi_0\ra\vert \leq C_R
  \Vert\beta^{(1)}-\beta^{(2)}\Vert_\infty(1+\vert Y\vert^2){\rm e}^{-\frac{\vert Y\vert^2}{C_R}}}.
  \eeq
  Now from estimate on the derivative  of $g(\beta^{(\theta)})$,  using (\ref{xy})  and (\ref{barg2}),  we get 
  the estimate (\ref{coker}). $\square$
  
  \vspace{0.5cm}
  
  \noi
  {\bf Proof of Proposition \ref{quantest2}}\\
  The estimate (\ref{quantlip1})   is a direct consequence  of Lemma \ref{unif}.\\
  We can get  (\ref{quantlip2}) from estimate (\ref{coker}) as follows.\\
  Let us introduce the space $L^{2,s}(\R^{2d}) =\{u\in L^2(\R^{2d}),\; \la X\ra^su(X)\in  L^2(\R^{2d})\}$ where $\la X\ra^s = (1+\vert X\vert^2)^{s/2}$.  Recall the useful  estimate : \\ $<X+Y>^{-2}\leq 2 <Y>^{-2}<X>^{2}$. 
  \\
  From  (\ref{coker})   we  can deduce   that the  linear  operator $\widetilde{\delta U}$ with  kernel 
  $\tilde K_{\delta U}$ is  continuous  from $L^{2,2}(\R^{2d}) $  into $L^{2}(\R^{2d}) $. 
   Let us consider the integral kernel
   ${\cal K}_2(X,Y) =  \tilde K_{\delta U}(X,Y)<Y>^{-2}$.  We have to prove  that ${\cal K}_2(X,Y)$ is the kernel  of  a  bound operator ${\cal T}_{{\cal K}_2}$ in $L^2(R^{2d})$.  Denote
\beq\label{C1}
   M_{{\cal K}_2}  = \max\left\{\sup_X\int\vert{\cal K}_2(X,Y\vert dY, \sup_Y\int\vert{\cal K}_2(X,Y\vert dX\right\},
\eeq
   We have the well known $L^2$-norm  estimate 
   \beq\label{C2}
   \Vert{\cal T}_{{\cal K}_2}\Vert \leq M_{{\cal K}_2}  
   \eeq
   Then using (\ref{coker}) and (\ref{C2}) we   get  that $\widetilde{\delta U}$ is continuous   from $L^{2,2}(\R^{2d})$
    in $L^2(\R^{2d})$, with a norm estimate\\
   %%%%%%%%%%%%%%%%%%%%%%%%
   Introduce the Fourier-Bargmann tranform: $\tilde\psi(X) = (2\pi)^{-d/2}\la\varphi_X, \psi\ra$  which is well defined
     for every $\psi\in{\cal S}'(\R^d)$. Recall that $\psi\mapsto \tilde\psi$ is an isometry  from
      $L^2(\R^d)$ into $L^2(\R^{2d})$   and that  $\varphi_X$ is an eigenvector for the creation  operators  ${\bf a}_j = \frac{1}{\sqrt 2}(x_j+\frac{\partial}{\partial x_j})$ with eigenvalue 
     $ \alpha_j=\frac{q_j+ip_j}{\sqrt 2}$  if $X=(q,p)\in\R^d\times\R^d$.\\  Then for every $k\geq 0$ there   exists $C_k$  such  that 
     $$
     \Vert\tilde\psi\Vert_{L^{2,k}(\R^{2d})} \leq  C_k\Vert\psi\Vert_{{\cal H}^k(\R^d)}
     $$
     So we get, under the conditions of (\ref{delco}), 
     $$
     \Vert \delta U\psi\Vert_{L^2(\R^d)} \leq C_R\Vert \beta^{(1)}-\beta^{(2)}\Vert_\infty\Vert\psi\Vert_{{\cal H}^2(\R^d)}.
     $$
  This proves (\ref{quantlip2}). $\square$
  
  Finally  we have  proved Corollary \ref{quadquant}  which is the main result  of this section.
  
  \section{Application to Strichartz estimate  and NLS}
  In this section  we give a proof for Theorem \ref{main3}  and Theorem \ref{main4}.\\
  A motivation for   studying  linear quantum  dynamics with noise  is the get results  for non linear  Schr\"odinger equations. For that it is now well known that Strichartz inequality is very useful. For quadratic Hamiltonian   with  noise $\beta$  this inequality  is derived    from the a   Mehler-Van Vleck formula  for  the  Schwartz kernel of the propagator $U_{H_\beta}(t,s)$
   for $0<\vert t-s\vert<T$, with $T$ small enough.  
  \subsection{A local dispersive estimate}   
  We start with an almost explicit expression for the kernel of the propagator valid  with noise.
 \begin{proposition}\label{kern1} If the Hypothesis \ref{MVV1}     and  Hypothesis \ref{MV2}   are satisfied 
     then for  every $R>0$ there  exists $T_R>0$  such that for  every $t,s\in ]t_0, t_0+T_R]$
      and every $\beta$ such that $\Vert \beta\Vert_{\cal C^\mu}\leq R$
     the Schwartz kernel ${\mathcal K}_\beta(t,s)$ of   $U_{H_\beta}(t,s)$   is a $C^\infty$  function of $(x,y)$  given by the following formula
     \beq\label{MVV}
     {\mathcal K}_\beta(t,s;x,y) = (2i\pi)^{-d/2}{\rm det}^{-1/2}\big(B_\beta(t,s)\big){\rm e}^{iS_\beta(t,s;x,y)}
    \eeq
      where $S_\beta(t,s;x,y)$  is the classical action  along  the unique  classical  trajectory  joining
      $y$ to $x$ at time $s$.\\
       In particular  there exists $\gamma >0$ such that 
       $\det B_\beta(t,s) \geq \gamma\vert t-s\vert^d$  for  every $t\in{I_T}_R$. \\
    Let  $p_\beta(t,s;x,y)$  be the momentum of  the trajectory 
    $(q_s,p_s) = \Phi_{H_\beta}(t,s)(q,p)$.  Then we have : 
     \beq
     S_\beta(t,s;x,y) = \int_{s}^t\big(\dot q_u\cdot p_u - H_\beta(u, q_u,p_u)\big)du,\;\;
    {\rm  where }\;\; p=p(t,s;x,y).
     \eeq
  \end{proposition}
  {\bf Proof.}  The computation is well  known, the new fact here  is that we need to control the validity of this  computation with  the noise term in $\beta$.\\
  First of all let us remark that  the action $S_\beta$ is continuous in $\beta$ for the ${\cal C}^0$ topology. \\
 To obtain this property  it is  enough to assume that $H_\beta(t)$ is quadratic. From Euler identity we have 
  $H_\beta(t) = \frac{1}{2}\left(q\cdot\partial_qH_\beta+p\cdot\partial_pH_\beta\right)$.  So using  the Hamilton equations :
 $ \dot q=\partial_pH_\beta$, $\dot p=-\partial_qH_\beta$  we have 
 $$
  S_\beta(t,s;x,y) = \frac{1}{2}\left(p_t\cdot q_t - p_s\cdot q_s\right).
  $$
  Recall that $(q_t, p_t) = \Phi_{H_\beta}(t,s)(q_s, p_s)$, so continuity  properties in $\beta$ for $S_\beta$ is a consequence of continuity  for the flow  $\Phi_{H_\beta}$.\\
  Now, we shall use  here   computations taken from \cite{biro}.\\
  Notice that in formula (\ref{kpol})  and (\ref{kquad})    the phases $\Psi_\Theta$  and $\Psi_{\Theta,2}$  are quadratic in $z$. 
  So the integral  is the integral  of a Gaussian  and we have to compute a Gaussian integral (a particular case  of the stationary theorem).  For this computation  we use  the simpler notations : $A=A_\beta(t,s)$ and the same for $B, C, D$.\\
   We choose  here in  (\ref{kquad})  the  complex matrices $\Theta(t) = \Gamma(t)$  where \\
    $\Gamma(t):=(C+iD)(A+iB)^{-1}$ (see section.3). 
   The matrix of the quadratic part of $\Psi_\Theta$ was computed in  \cite{biro}:
\beq\label{mathess}
  \partial^2_{z,z}\Psi = \begin{pmatrix}2i\1+(A+iB)^{-1}B & i(A+iB)^{-1}B\\  i(A+iB)^{-1}B & -(A+iB)^{-1}B    \end{pmatrix}   
  \eeq
  In particular  we get
\beq\label{detreg}
  {\rm det}\Big(\partial^2_{z,z}\Psi\Big)  = {\rm det}\Big(-2i(A+iB)^{-1}B\Big)
 \eeq
  and   $\partial^2_{z,z}\Psi$ is invertible if and only if $B:=B_\beta(t,s)$   is invertible. \\ This is checked   using the following lemma (in \cite[Proposition 2.30]{du} a similar result  is proved).
  \begin{lemma}\label{estflow}
  We have the following  estimate  of the flow $\Phi_{H_\beta}(t,s)$, for $\vert t-s\vert$  small enough,
    \bea\label{estflow1}
  \Phi_{H_\beta}(t,s) = \1 + ((\beta(t) -\beta(s))JS_K + (t-s)JS_H(s)  + \nonumber\\
  O\Big(\vert t-s\vert^2 + \sup_{\vert t-u\vert\leq\vert t-s\vert}\vert \beta(t)-\beta(u)\vert^2  \Big)
  \eea
  In particular if $\beta\in{\cal C}^\mu(I_T)$  with $\mu>\frac{1}{2}$  then we have
   \beq\label{estflow2}
  B_\beta(t,s) = (t-s)\partial^2_{pp}H(s) + O\big(\vert t-s\vert^2).
  \eeq
 Moreover if $\partial_{q,p}^2 K=0$  then the  estimate (\ref{estflow2}) remains true  for any $\beta\in{\cal C}_R^0({I_T}_R)$.\\
  In estimates (\ref{estflow1}) (\ref{estflow2})  the big O   is uniform  for $\Vert \beta\Vert_{\cal C^\mu}\leq R$.
  \end{lemma}
  Using lemma \ref{estflow} and choosing $T_R>0$  small enough we get that $B_\beta(t,t_0)$ is invertible for  
  $t\in{I_T}_R$.   So under  the same conditions we have   that
  $$
   {\rm det}\big(\partial^2_{z,z}\Psi (t,s)\big)\neq 0.
   $$
   So we get  (\ref{MVV})  by computing  a Gaussian integral.  \\
  {\bf Proof  of lemma \ref{estflow}}\\
  Using (\ref{hamint})  we get 
  \bea\label{iterap}
  \Phi_{H_\beta}(t,s) = {\rm e}^{(\beta_t-\beta_{s})JS_K} + \int_{s}^t{\rm e}^{(\beta_t-\beta_{u})JS_K}
  JS_H(u){\rm e}^{(\beta_u-\beta_{s})JS_K}du \nonumber\\
  + \int_{s}^t{\rm e}^{(\beta_t-\beta_{u})JS_K}JS_H(u)
  \Big(\int_{s}^s{\rm e}^{(\beta_u-\beta_{\sigma})JS_K}\Phi_{H_\beta}(\sigma, s)d\sigma\Big)du
  \eea
  The  last   term  is clearly $O(\vert t-s\vert^2)$. To estimate  the  first  we use  the  Taylor  formula
  \beq\label{tayl}
  {\rm e}^{uJS_K} = 1 + uJS_K + u^2(JS_K)^2\int_0^1(1-\theta){\rm e}^{\theta uJS_K}.
  \eeq
  Notice  that  we have $JS_K = \begin{pmatrix} L_K & 0\\ G_K & L_K^\top  \end{pmatrix}$
  and $JS_H(t) = \begin{pmatrix} L_(t) & E_H(t)\\ G_H(t) & L_H(t)^\top  \end{pmatrix}$.
  Notice that $E_H(t)$  is  invertible  for  $t$ close to $t_0$. Moreover $JS_K^2=0$ if 
  $ L_K = 0$. So  the lemma  can be easily obtained  from (\ref{iterap})  and (\ref{tayl}) . $\square$\\
  The next corollary  is very useful  in applications  to get Strichartz  estimates,  as  explained in \cite{keta}.
  
  \begin{corollary}[dispersive estimate]
  There  exists  a constant $C_R$, depending  only on $R$ such that  for  every $t\in{I_T}_R$   and every $x, y\in\R^d$, we have 
  \beq\label{unifest}
   \vert{\mathcal K}_\beta(t,s;x,y)\vert \leq C_R\vert t-s\vert^{-d/2}.
  \eeq
  and for every $p\in[2, +\infty]$,   we have  for $\psi\in {L^p(\R^d)}$, 
  \beq\label{stri}
  \Vert U_\beta(t,s)\psi\Vert_{L^p(\R^d)}  \leq 
  C_R\vert t-s\vert^{-d(1/2-1/p)}\Vert\psi\Vert_{L^{p'}(\R^d)},\; 1/p+1/p' = 1.
  \eeq
  \end{corollary}
  Let  us   notice that  the principal symbol of $\hat H(t)$ is not  necessary elliptic,
   the important property to  get the  local   dispersive estimate (\ref{stri}) 
    is that the quadratic form $\partial^2_{p,p}H$  is non degenerate (for $d=2$ we may have 
    $H(q,p) = p_1^2 -p_2^2 + q_1^2 + q_2^2$). 
  
  \subsection{About the proof of Theorem \ref{main4}}
 As already remarked in \cite{bofu, du}, using Strichartz estimate (\ref{stri}) it is possible  to extend the results  proved  in  \cite{ca}  concerning non  linear  Schr\"odinger equations for quadratic linear  parts  with noise. The proofs follows  closely \cite{ca}  so  we do not  repeat the detils here (see  also \cite{caz, lipo})  for the regular case).\\
In a first  step  the result is  proved locally in time by a fixed point  argument  such  that $\beta\mapsto\psi_\beta(t)$ is continuous from 
   ${\cal C}^\mu(I_T)$ into $L^2(\R^d)$. Then we get  the conservation  of the $L^2$ norm (this is true for $\beta\in C^1(I_T)$   and also for $\beta\in {\cal C}^\mu(I_T)$ by continuity).
   Using  the conservation law   we  can extend the  local  solution  in a global  solution  for initial data in $L^2(\R^d)$  for subcritical non linearities $\sigma$.

    The  second part of  Theorem  \ref{main4} gives a local  well-posedness result  in ${\cal H}^1(\R^d)$ and can be proved following   closely  \cite[Proposition 2.5]{ca}  as a consequence  of  the   dispersive  estimate (\ref{stri}) for $\beta\in{\cal C}^0(I_T)$.

 %   \begin{theorem} Assume that  the conditions (MVV1)  and (MVV2) are  satisfied  and that one of the following conditions are  satisfied:\\
   % (i) $\sigma<\frac{\sigma}{2}$ (the non linearity is $L^2$-subcritical)\\
 %   (ii) $\frac{2}{d} \leq \sigma <\frac{2}{d-2}$, $d\geq 2$  and $\lambda \geq 0$ (defocusing case)\\
   %Then  for any initial data $\psi(s)\in {\cal H}^1(\R^d)$, the  integral equation (\ref{NLSI}) has a unique solution 
%$\psi_\beta\in {\mathcal C}^0(\R,{\cal H}^1(\R^d))$.
%  \end{theorem}
  %%%%%%%%%%%%%%%%%
%%%%%%%%%%%%%%%%%%
%\section{Perturbations of sub-quadratic Hamiltonians}

%\section{Perturbations  by   several  noise  sources}

%%%%%%%%%%%%%%%%%%%%%%%%%%%%%%%%%%%%%%%%%

\rule[0pt]{\hsize}{0.4pt}

\noi
D\'epartement de Math\'ematiques,  Laboratoire Jean Leray, CNRS-UMR 6629\\
Universit\'e de Nantes, 2 rue de la Houssini\`ere,  F-44322 NANTES Cedex 03, France\\
E-mail adress: \texttt{didier.robert@univ-nantes.fr}

\end{document}